\magnification\magstep0
\hsize = 12.0 cm
\vsize = 18.6 cm
\overfullrule 0pt
%
%
\nopagenumbers
\tenrm

\noindent{\bf Hard Simulation Problems in the Modeling of 
Magnetic Materials: Parallelization and Langevin Micromagnetics}
\vskip  0.15 true cm
\noindent{G.~Korniss$^{1}$, 
          G.~Brown$^{1,2}$, 
          M.~A.~Novotny$^{1}$, and  
          P.~A.~Rikvold$^{1,2}$
\vskip  0.15 true cm
\noindent{$^1$Supercomputer Computations Research Institute, 
          Florida State University, Tallahassee, Florida 32306-4130 }
\vskip -0.05 true cm
\noindent{$^2$Center for Materials Research and Technology 
          and Department of Physics,
          Florida State University, Tallahassee, Florida 32306-3016}

\vskip 0.5 true cm

\noindent{\bf Abstract.}  
We present recent results on two attempts at understanding 
and utilizing large-scale simulations of magnetic materials.  
In the first study 
we consider massively parallel implementations on a Cray T3E 
of the $n$-fold way algorithm 
for magnetization switching in kinetic Ising models.  
We find an intricate 
relationship between the average time increment and the size of the 
spin blocks on each processor.  
This narrows the regime of efficient implementation.
The second study concerns incorporating noise into micromagnetic 
calculations using Langevin methods.  
This allows measurement of quantities such as 
the probability 
that the system has not switched within a given 
time.  Preliminary results are reported for 
arrays of single-domain nanoscale pillars.  

\vskip 0.2 true cm

\noindent{\bf 1.~Introduction}
\vskip 0.1 true cm
\noindent  
To model realistic magnetic systems of interest in, for example, 
the magnetic recording industry requires that a number of difficult 
simulational problems be addressed.  Preliminary results on 
two such problems are presented here.  

Simulating metastable decay involves long characteristic time scales
(the metastable lifetime), and several sophisticated algorithms 
have been developed for serial computers [1,2,3].  
A common testbed for these algorithms is the kinetic Ising 
ferromagnet below its critical temperature, $T_c$, which exhibits slow 
metastable decay after the reversal of the external magnetic field [4].  
This model is appropriate for the study of highly anisotropic 
single-domain nanoparticles and thin films [5].  
Even with these sophisticated 
algorithms the computer power required is enormous.  
Efficient computation 
requires these algorithms to be 
scalable and effectively implemented on massively parallel computers.  
We present an experiment on the parallelization of $n$-fold algorithms.  

For less anisotropic magnetic materials, 
continuous-spin models should be simulated.  To model 
metastable decay at finite temperature and measure time
dependent quantities of experimental interest requires 
extensions of normal micromagnetic calculations.  We report 
our first Langevin micromagnetic calculation, 
designed to model arrays of pillars grown with 
an STM technique [6].  

\vskip 0.25 truecm
\noindent{\bf 2.~Parallelization of the $n$-fold way algorithm}
\vskip 0.1 true cm
\noindent  
We present and analyze a variation 
[7] of the 
$n$-fold way algorithm 
[1,2] for magnetization switching in 
the kinetic Ising model on a distributed-memory parallel computer. 
The implementation of efficient massively parallel algorithms for Monte 
Carlo simulations is an interesting and challenging problem, which is 
one of the most complex ones in parallel computing. It belongs to the class 
of parallel discrete-event simulation (sometimes referred to as
distributed simulation) which has numerous applications in engineering, 
computer science, and economics, as well as in physics [8]. 
These dynamics, which obviously contain a substantial 
amount of parallelism, were traditionally simulated on serial computers. 
Paradoxically, it is difficult to implement an efficient 
parallel algorithm to simulate them, mainly due to the fact that the 
discrete events are not synchronized by a global clock.

The kinetic Ising model, either with the standard integer-time updates or with
Glauber's continuous-time interpretation, was believed to be 
inherently serial.  
Contrary to that belief, Lubachevsky presented a method for
parallel simulation of these systems 
[7] {\it without} changing the 
underlying dynamics.  
Also, he proposed a way to incorporate the 
$n$-fold way algorithm, possibly giving further speedup. We implement
his algorithm on the isotropic, square-lattice Ising model 
with periodic boundary conditions and 
Hamiltonian
{${\cal H}$$=$$-J \sum_{\langle i j \rangle} s_i s_j -H \sum_i s_i$}.  
Here $J$$>$$0$ is the 
ferromagnetic nearest-neighbor spin-spin interaction and
$H$ is the external field. To study metastable decay, all spins are
initialized in the $+1$ state, and we apply a negative magnetic field at 
constant $T<T_c$. In the corresponding serial algorithm we use the 
single-spin-flip Metropolis rates, where the probability to flip a spin is
$p$$=$$\min\{1,\exp(-\Delta {\cal H})\}$. 
In the rejection-free $n$-fold way update scheme, a 
flip is always performed, 
and the time is incremented appropriately.
One must then introduce the notion of spin classes which carry the
state of the spin itself and its neighbors. In the above model 
there are ten classes, characterized by the number
of spins in class $i$, $n_i$, and the flipping probability 
of a spin in class $i$, $p_i$.  
Since the classes are disjoint, 
$\sum_{i=1}^{10} n_i$$=$$L^2$, where $L$ is the linear size of the 
lattice. 
To perform an update, first a class is chosen according to the
relative weights $\{n_i p_i\}_{i=1}^{10}$, then one of the spins in the class
is picked with equal probability, $1/n_i$.  Once the class information, 
in particular the $n_i$'s, have been updated, the time of the next 
update is determined. The time increment is a random quantity, 
given by $-\ln(r) L^{2}/\sum_{i=1}^{10} n_i p_i$, where $r$ is a uniformly 
distributed random number in $(0,1]$. 
For integer-time updates, the only difference is that one must
draw the time increments from a discrete geometrical distribution instead of
the continuous exponential one [2].

To parallelize the above algorithm, 
the $L$$\times$$L$ lattice is 
spatially decomposed
into $(L/l)^2$ blocks of size $l$$\times$$l$.  
On a parallel computer, each processing element 
(PE) carries an $l$$\times$$l$ block of spins and the number of PEs is 
$N_{PE}$$=$$(L/l)^2$. 
However, one cannot simply run a copy of the serial $n$-fold way algorithm on 
each PE without the possibility of corrupting the history of neighboring PEs.  
On each PE 
an additional class is defined which contains the spins on its boundary. The
relative weight of this class is the number of spins on the 
boundary, $N_b$$=$$4(l-1)$, 
which clearly does not change during the simulation.
The original tabulation 
of spins is only used in the kernel of the block.
Hence, $N_b+\sum_{i=1}^{10} n_i$$=$$N$, where $N$$=$$l^2$ 
is the total number of 
spins in a block. The update scheme differs from the 
original (continuous-time) algorithm in the following steps: 
{\it (i)}  if the spin chosen belongs to the boundary, then the updating PE 
must {\it wait until} its local time becomes less or equal than that of its 
neighboring PEs (at most two in two dimensions). Then the state of this spin 
{\it may or may not} change: its flipping probability is determined by the 
usual Metropolis rates. 
{\it (ii)} once an update is completed, the time of the next update is 
determined by the {\it local} time increment,
$$
\Delta t=-N\ln(r) \left/ \left[N_b + \sum_{i=1}^{10} n_i p_i\right] \right. 
\;.
\eqno{(1)} 
$$

\null
\vskip 5.6 truecm 
\vskip -10 truept
\includegraphics{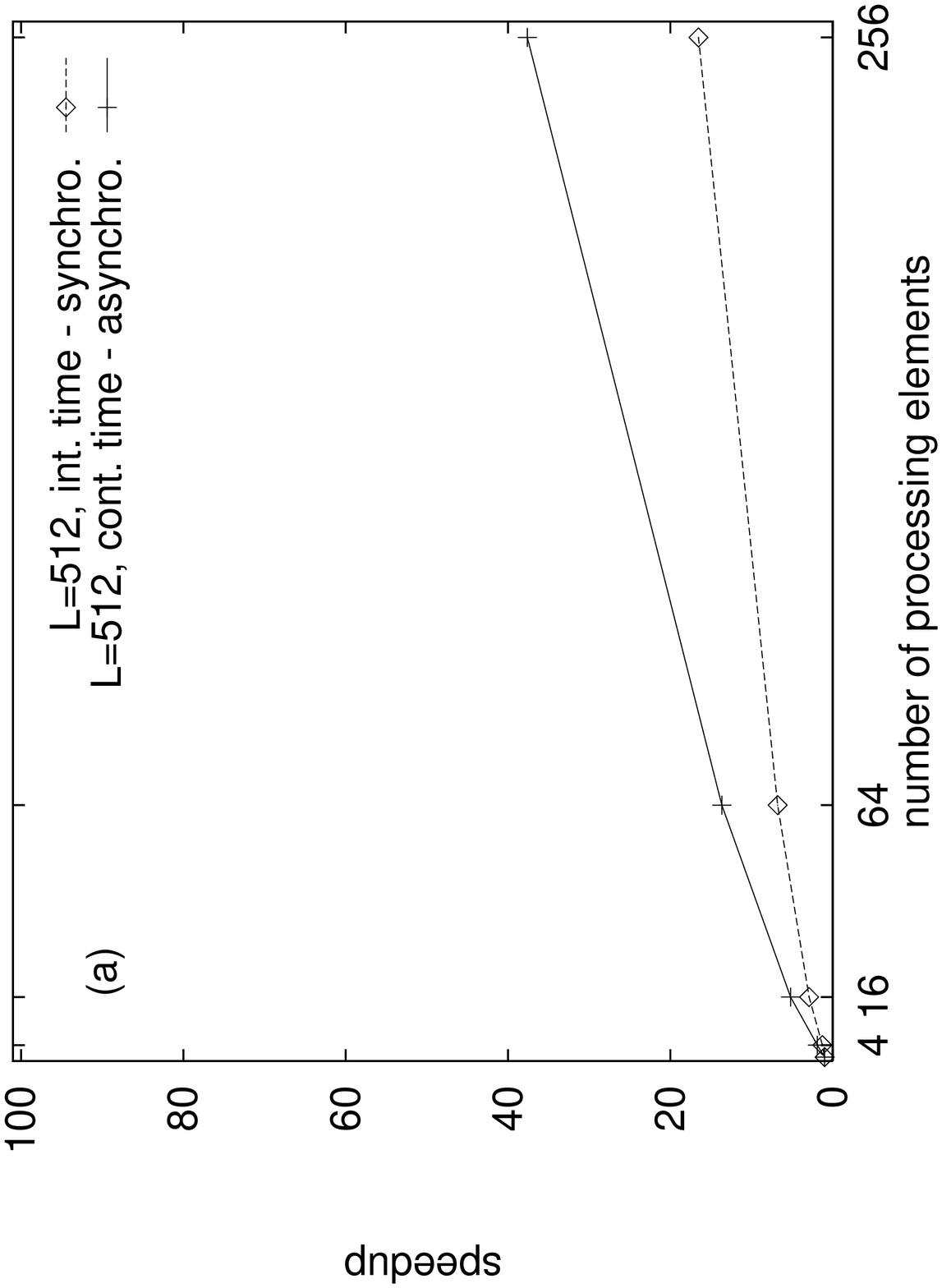} 
\includegraphics{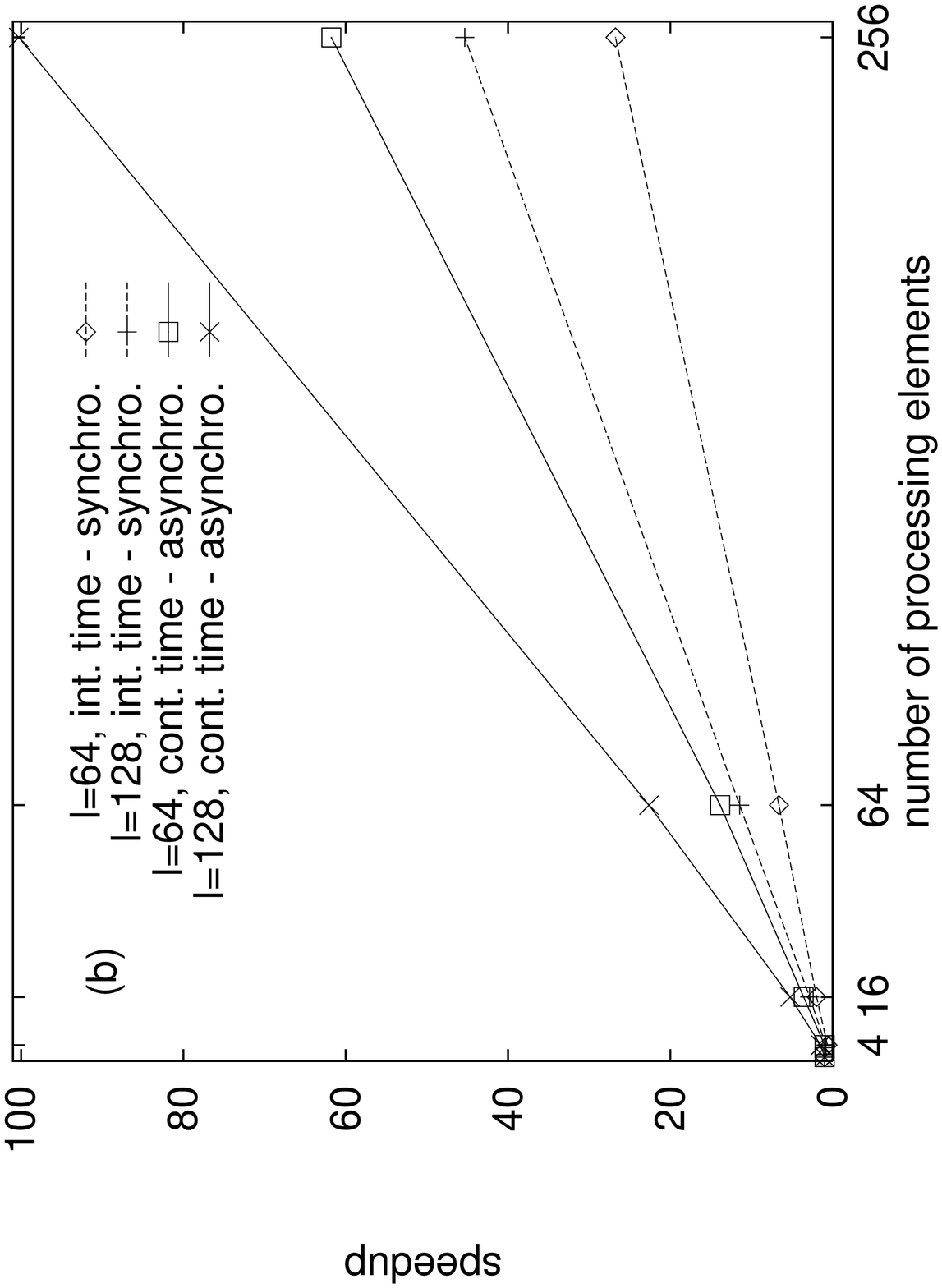} 

\noindent {\bf Fig.~1}~
Speedup measurements for the parallel code as a function of the
number of processing elements, $N_{\rm PE}$.
(a) For fixed system size, $L$$=$$512$, the block size, $l$, decreases 
with increasing $N_{\rm PE}$.
(b) For fixed block size, $l$$=$$64$ and $128$, the system size $L$ is 
increasing with increasing $N_{\rm PE}$.

\vskip 0.25 truecm

It is clear from the above algorithm that at any given (wall clock) moment
different PEs generally have different local simulated times. The
``{\it wait until\/}'' control structure in {\it (i)}, 
however, ensures that the 
information passed between neighboring PEs does not violate causality [7].  
The above asynchronous algorithm is suitable for a continuous-time
update scheme, but it can cause inconsistency when integer time is used.
Then, to ensure the reproducibility of a simulated path, provided the
same set of random seeds are used, explicit barrier synchronization should be 
incorporated (synchronous algorithm).

We implement the above versions of the $n$-fold way algorithm on the
Cray T3E parallel computer at NERSC, using the Cray-specific, 
logically shared, distributed memory access (SHMEM) routines for message 
passing. The fast SHMEM library supports communication initiated by {\it one} 
PE, together with remote atomic memory operations. Without these features, it 
would be extremely inconvenient to code an algorithm for stochastic simulation
on a distributed memory machine, where the communication pattern is 
completely unpredictable. These characteristics outweigh the loss of 
portability of our code. Details on the implementation will be published 
elsewhere [9].

We note some inherently
weak features, which are not related to the fast communication
hardware of the parallel architecture.  
First, as a general guideline, the fewer communications needed, 
the better the performance of the parallel code. In our case, the probability 
to pick a spin on the boundary, which is ultimately 
followed by some communication, is greater than the surface-to-volume
ratio. It is determined by the relative weights in the 
modified $n$-fold way algorithm, 
$N_{b}/(N_{b}+\sum_{i=1}^{10} n_i p_i)$. With very small $p_i$'s this ratio
can become close to $1$, leading to more frequent 
message passing and idling as required by 
the ``{\it wait until\/}'' condition. 
Second, the average time increment in Eq.~(1) 
is not bounded by
$p_{min}^{-1}$ as in the serial $n$-fold way algorithm, but by
$$
\langle\Delta t\rangle _{max}={{1}\over{N_{b}}/N + p_{min} N_{k}/N} <
{{l^2}\over{4(l-1)}}\;,
\eqno{(2)}
$$
where $N_k$ is the number of spins in the kernel.  Hence, 
however small the flipping probabilities, the average local
time increment is limited by approximately $l/4$.  
Consequently, reasonable performance requires 
$l p_{min} {>}{\sim}1$. 

We tested the scaling of the code (both asynchronous and synchronous 
versions) up to 256 PEs at $T$$=$$0.7T_c$ and $|H|/T$$=$$0.18$, 
in two different
ways. First, the system size is kept constant ($L$$=$$512$), 
and we divide it into 
smaller and smaller blocks (Fig.~1a). Second, 
we keep the block size fixed ($l$$=$$64, 128$), and study larger 
systems by increasing the number of blocks (Fig.~1b).
We determine the efficiency and speedup by
comparing with the serial $n$-fold way performed
on one T3E node.  
The results reflect the features discussed in the previous paragraph.
In the first case we observe poor scaling, due to drastically decreasing average
time increments and a slightly decreasing utilization ratio.  In the second
case, the average time increments are not affected while the utilization
saturates, leading to reasonably good scaling. The larger the block size, 
the better the performance. For the continuous time, asynchronous
algorithm, with  $l$$=$$128$ 
and using $256$ PEs, the speedup is $100$.  It can 
be systematically improved by taking larger $l$ values. 
However, the memory of a PE is not unlimited: the largest
cell size that we could allocate in a T3E node was $l$$=$$1400$. 
The asynchronous algorithm suits this distributed-memory architecture best.  

The practical applicability of our implementation is obviously driven
to large systems. 
The narrow regime of efficient implementation is due to 
the introduction of a special class in the $n$-fold way algorithm which 
``shields'' the blocks from each other. The algorithm avoids rollbacks, but
pays a large price: it looses the most important feature of the serial
$n$-fold way algorithm --- the arbitrarily large time increments at 
arbitrarily low temperature and field. 
The only way to preserve the advantage 
of the original $n$-fold way algorithm is to apply it directly on each 
block, together with a complex rollback procedure [10].  
Work is in progress to incorporate
it in our simulations of metastable decay.

\noindent  
\null
\vskip 7 truecm 
\vskip -27 truept
\includegraphics{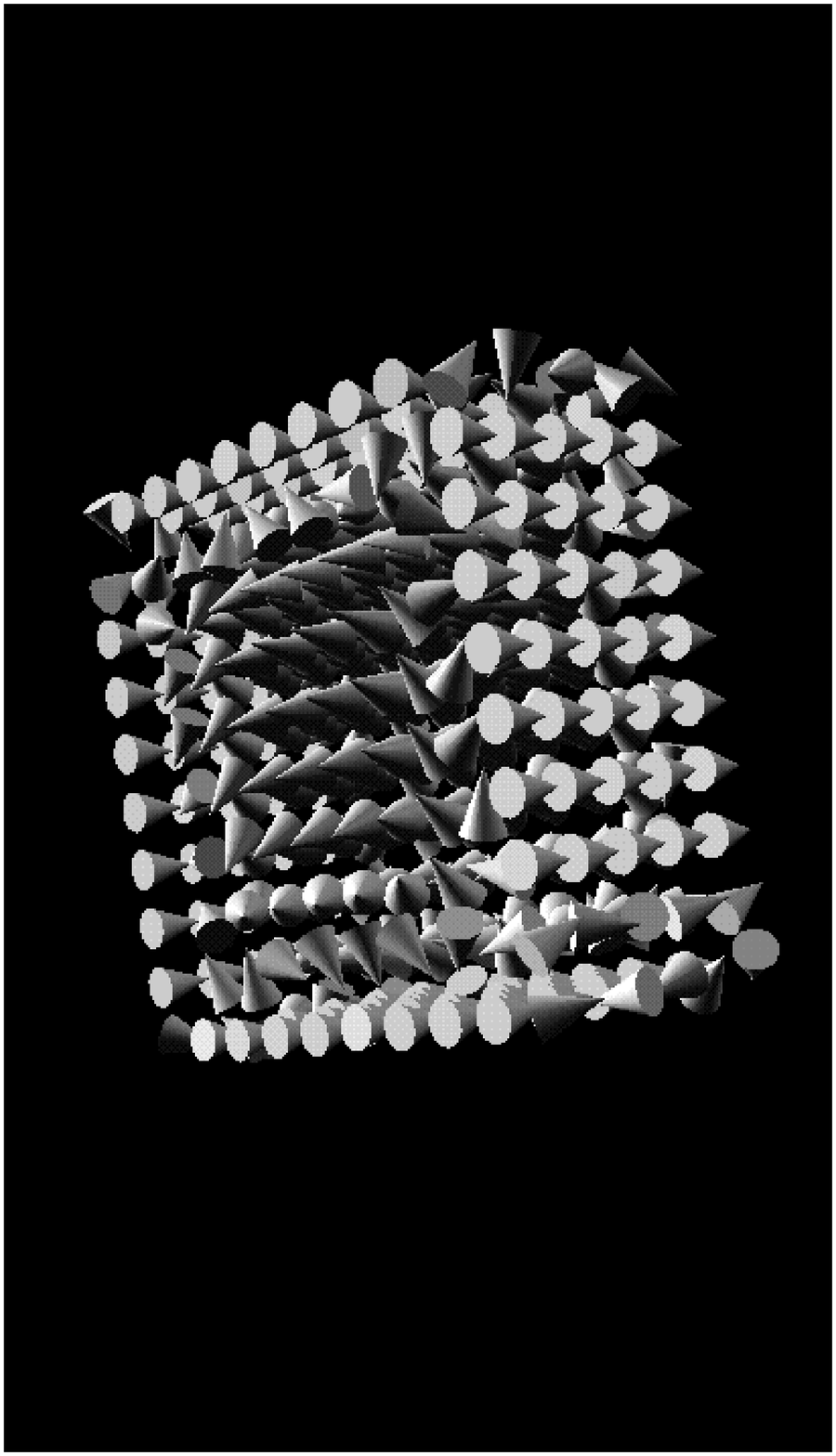} 

{\bf Fig.~2}~
A single snap-shot of a 
Langevin micromagnetic calculation for magnetization 
reversal in a square array of 
Ni pillars that are 200~nm apart, 200~nm tall, and have a 
diameter of 40~nm.  
Each pillar is discretized using 5 lattice points.  
(The vertical scale of this figure 
is enhanced compared with the horizontal scale for 
clarity of presentation.)  
The temperature is 300~K, the spins are initially 
up, and the 
applied field is down with a magnitude of $1225$~Oe.  
This is at a time of 14~nsec following the reversal of the field.  
The integration time step is $\Delta t$$=$1~psec.  
Note that this looks very different from a coherent rotation mode 
of spin reversal.  

\vskip 0.15 truecm

\noindent{\bf 3.~Langevin Micromagnetics}
\vskip 0.1 true cm
\noindent 
In order to simulate systems for which the Ising model is not a 
faithful representation, we have programmed a Langevin micromagnetics 
code similar to that reported in [11].  
With a phenomenological damping parameter $\alpha$, 
and classical spins of 
constant length given by the bulk saturation magnetization $M_{\rm s}$, 
at each lattice site $i$ we have a scaled magnetization 
${\vec m}={\vec M}_{\rm s}/M_{\rm s}$.  
The standard Ginzburg-Landau-Lifshitz micromagnetic equation [12,13] is 
$$
{{d{\vec m}_i}\over{dt}} = 
-{1\over{1+\alpha^2}} 
{\vec m}_i\times
\left({\vec h}_{i,{\rm eff}}+\alpha{\vec m}_i\times
{\vec h}_{i,{\rm eff}}\right)
\; .
\eqno(3)
$$
The scaled field at each site, ${\vec h}_{i,{\rm eff}}$, 
contains contributions from terms including 
the dipole-dipole interaction, 
the exchange interaction, 
the interaction due to crystaline anisotropy, 
the applied field, 
and 
a scaled noise term proportional to the 
the Langevin fields $\zeta(t)$ [12,13].  
In our case the Langevin noise term $\zeta$ and 
the integration time step $\Delta t$ 
are related by $\zeta$$\propto$$\sqrt{\Delta t}$, which we think 
is more physical than the 
$\zeta$$\propto$$1$$/$$\sqrt{\Delta t}$ of [11].  
Even though the set of equations used in this Langevin micromagnets 
simulation are approximations to the actual equations [14], 
the approximation should be reasonable well below 
the critical temperature.  
In order to keep the length of the ${\vec m}_i$ constant, we have used a 
fourth-order Runge-Kutta algorithm.  
Fig.~2 shows the type of simulations [15] 
that can be performed for arrays of magnetic pillars 
that can be built and measured experimentally [6].  
The importance of the finite temperature thermal fluctuations 
and the rotation mode, which is very different from that of 
uniform rotation, can be seen in this figure.  
Note that with standard time-independent micromagnetic calculations 
there would be no magnetization reversal, 
since the strength of the applied field is smaller than that of 
the nucleation field.  


\vskip 0.25 truecm
\noindent{\bf Acknowledgements\/}.  
Special thanks to M.~Kolesik for invaluable discussions and to R.~Gerber
at the NERSC consulting group for helpful hints 
on debugging the parallel code.  
This research was supported by 
NSF Grant No.\ DMR-9520325, 
FSU-SCRI (DOE Contract No. DE-FC05-85ER25000), 
computer time allocated on the NERSC T3E by the DOE, 
and 
FSU-MARTECH.  

\vskip 0.15 truecm

\noindent{\bf References}
\vskip 0.1 truecm

\noindent [1] A.B.\ Bortz, M.H.\ Kalos, J.L.\ Lebowitz, 
J.\ Comput.\ Phys.\ {\bf 17}, 10 (1975).
\vskip 0.0 true cm

\noindent [2] M.A.\ Novotny, 
Phys.\ Rev.\ Lett.\ {\bf 74}, 1 (1995);
erratum {\bf 75\/}, 1424 (1995); 
\vskip 0.0 true cm
\noindent ~~~~~~Comp.\ in Phys.\ {\bf 9}, 46 (1995).
\vskip 0.0 true cm

\noindent [3] 
M.\ Kolesik, M.A.\ Novotny, P.A.\ Rikvold, and D.M.\ Townsley,
in {\it Computer\/} 
\vskip 0.0 true cm
\noindent ~~~~~~{\it Simulations in Condensed Matter Physics~X},
edited by D.P.\ Landau, K.K.\ 
\vskip 0.0 true cm
\noindent ~~~~~~Mon, and H.-B.\ Sch\"uttler,
(Springer, Berlin, 1998), pp.~246; 
M.\ Kolesik, 
\vskip 0.0 true cm
\noindent ~~~~~~M.A.\ Novotny, and P.A.\ Rikvold, 
Phys.\ Rev.\ Lett., in press.  
\vskip 0.0 true cm

\noindent [4]~P.A.\ Rikvold, H.\ Tomita, S.\ Miyashita and S.W.\ Sides,
Phys.\ Rev.\ E 
\vskip 0.0 true cm
\noindent ~~~~~~{\bf  49}, 5080 (1994).
\vskip 0.0 true cm

\noindent [5] H.L.~Richards, S.W.~Sides, M.A.~Novotny, and P.A.~Rikvold,
J.\ Mag.\ Mag.\ 
\vskip 0.0 true cm
\noindent ~~~~~~Mat.\ {\bf 150}, 37 (1995); 
P.A.\ Rikvold, M.A.\ Novotny, M.\ Kolesik, and H.L.\ 
\vskip 0.0 true cm
\noindent ~~~~~~Richards,
in {\it Dynamical Properties of Unconventional Magnetic Systems\/}
\vskip 0.0 true cm
\noindent ~~~~~~edited by A.T.\ Skjeltorp and D.~Sherrington,
(Kluwer, Dordrecht, 
\vskip 0.0 true cm
\noindent ~~~~~~in press) NATO ASI Series; 
M.~Kolesik, M.A.\ Novotny, and P.A.\ 
\vskip 0.0 true cm
\noindent ~~~~~~Rikvold, Phys.\ Rev.\ B {\bf 56}, 11790 (1997).
\vskip 0.0 true cm

\noindent [6] 
A.D.\ Kent, S.\ von Moln{\'a}r, S.\ Gider, and D.D.\ Awschalom, 
J.\ Appl.\ Phys.\
\vskip 0.0 true cm
\noindent ~~~~~~{\bf 76},  6656  (1994); 
S.~Wirth, M.~Field, D.D.\ Awschalom, and S.\ von~Moln{\'a}r, 
\vskip 0.0 true cm
\noindent ~~~~~~Phys.\ Rev.\ B, submitted.  
\vskip 0.0 true cm

\noindent [7] B.D.\ Lubachevsky, Complex Systems {\bf 1}, 1099 (1987);
\vskip 0.0 true cm
\noindent ~~~~~~J.\ Comput.\ Phys.\ {\bf 75}, 103 (1988).
\vskip 0.0 true cm

\noindent [8] 
R.M.\ Fujimoto, Commun.\ ACM {\bf 33}, 30 (1990).
\vskip 0.0 true cm

\noindent [9] 
G.\ Korniss, M.A.\ Novotny, and P.A.\ Rikvold, in preparation. 
\vskip 0.0 true cm

\noindent [10] 
D.R.\ Jefferson, 
ACM Trans.\ Prog.\ Lang.\ and Syst.\ {\bf 7}, 404 (1985).  
\vskip 0.0 true cm

\noindent [11] E.D.\ Boerner and H.N.\ Bertram, 
IEEE Trans.\ Magn.\ {\bf 33}, 3052 (1997).
\vskip 0.0 true cm

\noindent [12] W.F.\ Brown, Phys.\ Rev.\ {\bf 130}, 1677  (1963).
\vskip 0.0 true cm

\noindent [13] A.~Aharoni, 
{\it Introduction to the Theory of Ferromagnetism\/} 
(Clarendon, 
\vskip 0.0 true cm
\noindent ~~~~~~Oxford, 1996).  
\vskip 0.0 true cm

\noindent [14] D.A.\ Garanin, Phys.\ Rev.\ B {\bf 55}, 3050 (1997).
\vskip 0.0 true cm

\noindent [15] G.\ Brown, M.A.\ Novotny, and P.A.\ Rikvold, 
in preparation.  

\end